\documentclass[12pt, JPCM]{iopart}
\usepackage{amssymb}
\usepackage{iopams}
\usepackage{hyperref}
\usepackage{graphicx}
\usepackage[numbers,sort&compress]{natbib}
\usepackage{color}

\hypersetup{hypertex=true,
colorlinks=true,
linkcolor=blue,
urlcolor = blue,
citecolor=blue}

\begin{document}

\title{Exceptional spectrum and dynamic magnetization}
\author{Y B Shi, K L Zhang and Z Song*}
\address{School of Physics, Nankai University, Tianjin 300071, China}
\begin{indented}
	\item[]*Author to whom any correspondence should be addressed.
\end{indented}
\ead{songtc@nankai.edu.cn}
\vspace{10pt}
\begin{indented}
	\item[]\today
\end{indented}

\begin{abstract}
A macroscopic effect can be induced by a local non-Hermitian term in a
many-body system, when it manifests simultaneously level coalescence of a
full real degeneracy spectrum, leading to exceptional spectrum. In this
paper, we propose a family of systems that support such an intriguing
property. It is generally consisted of two arbitrary identical Hermitian
sub-lattices in association with unidirectional couplings between them. We
show exactly that all single-particle eigenstates coalesce in pairs even
only single unidirectional coupling appears. It means that all
possible initial states obey the exceptional dynamics, resulting in some
macroscopic phenomena, which never appears in a Hermitian system. As an
application, we study the dynamic magnetization induced by complex fields in
an itinerant electron system. It shows that an initial saturated
ferromagnetic state at half-filling can be driven into its opposite state
according to the dynamics of high-order exceptional point. Any
Hermitian quench term cannot realize a steady opposite saturated
ferromagnetic state. Numerical simulations for the dynamical processes of
magnetization are performed for several representative situations, including
lattice dimensions, global random and local impurity distributions. It shows
that the dynamic magnetization processes exhibit universal behavior.
\end{abstract}

\maketitle

\section{Introduction}

A local Hermitian magnetic field cannot induce a global
magnetization. It is well known that a non-Hermitian system may make many
things possible, based on development of non-Hermitian quantum mechanics,
both in theoretical and experimental aspects~\cite%
{PRL08a,PRL08b,Klaiman,CERuter,YDChong,Regensburger,LFeng,Fleury,CMBender,NM,FL,Ganainy18,YFChen,Christodoulides}%
. These include quantum phase transition that induces in a finite system~ 
\cite%
{Znojil1,Znojil2,Bendix,LonghiPRL,LonghiPRB1,Jin1,Znojil3,LonghiPRB2,LonghiPRB3,Jin2,Joglekar1,Znojil4,Znojil5,Zhong,Drissi,Joglekar2,Scott1,Joglekar3,Scott2,Tony}%
, unidirectional propagation and anomalous transport~\cite%
{LonghiPRL,Kulishov,LonghiOL,Lin,Regensburger,Eichelkraut,LFeng,Peng,Chang},
invisible defects~\cite{LonghiPRA2010,Della,ZXZ1,Jin4}, coherent absorption 
\cite{Sun} and self sustained emission~\cite%
{Mostafazadeh,LonghiSUS,ZXZSUS,Longhi2015,LXQ,Jin3}, loss-induced revival of
lasing~\cite{PengScience}, as well as laser-mode selection \cite%
{FengScience,Hodaei,JLPRL}. Such kinds of novel phenomena can be traced to
the existence of exceptional point (EP), which is a transition point of
symmetry breaking for a pair of energy levels. It occurs when eigenstates
coalesce~\cite{CMBender,NM,AAlu}, and usually associates with the
non-Hermitian phase transition~\cite{LFeng,YFChen}. The EP has many
applications in optics~\cite%
{Klaiman,Doppler,Xu,Assawaworrarit,Midya,HouZL,Alu}, not only involving
non-reciprocal energy transfer \cite{Xu}, but also unidirectional lasing~%
\cite{PMiao16,Longhi17}, and optical sensing~\cite{WChen,Hodaei17}.

A fundamental question is whether a single impurity can induce multiple-EP,
resulting macroscopic quantum phenomena in a many-body system. 
It is possible according to the conclusion in reference \cite{WPPRB}. The
key point is how to construct such a system. Considering an extreme but
simplest case, $2N$ non-degeneracy energy levels become $N$ energy levels by
pairing coalescing. We dub the resulting set of energy levels as coalescing
spectrum. In this situation, every fermion obeys exceptional
dynamics simultaneously for half-filling case. The dynamics of an $N$%
-fermion state naturally results in a macroscopic effect in the
thermodynamic limit. So that the manifestation of macroscopic effect is not
surprising. In recent work, it has been shown that a continuous change of an
asymmetric hopping strength can induce a sudden change of the
single-particle spectral statistics, which results in non-analytic behavior
of some macroscopic quantities~\cite{WP}. In addition ~\cite{ZXZ2}, an EP
with the order of the size of the system is created by a single impurity in
an $N$-site quantum spin chain. Motivated by these works, we investigate the
mechanism of the appearance of single-particle coalescing spectrum
(high-order EP~\cite{ZSM,HH,JS,KD,Jin5} for many-particle spectrum) induced
by a single non-Hermitian impurity.

The purpose of the present work is to present a general formalism for a
coalescing spectrum, or a system of which every eigenstate is a coalescing
state. We study the possible structure of such a system, the
corresponding dynamics, and applications in physics. In contrast, most of
previous work focus on the cases with finite number of (or a portion of)
coalescing energy levels embedded into the real or complex spectrum. To
this end, we propose a family of systems which consists of two arbitrary
identical Hermitian sublattice in association with arbitrary unidirectional
couplings between them. Exact analysis shows that all single-particle
eigenstates coalesce in pairs, even an arbitrary unidirectional coupling
appears. This provides way to design an $2N$-site lattice system to possess $%
N$ energy levels. Accordingly, such a special spectral structure supports
intriguing dynamical behaviors which can not be achieved in the
framework of conventional quantum mechanics. As an application, we study
tight-binding model for spin-$1/2$ fermions at half-filling, with various
impurity distributions, which arise from global or local complex fields.
Numerical simulation shows that the dynamic magnetization processes exhibit
universal behavior. Even a local complex field can drive a global
magnetization due to the EP-related dynamics.

The remainder of the paper is organized as follows. In section \ref{II}, we
present a class of system which possesses exceptional spectrum. In
subsection \ref{IIa}, we map the non-Hermitian Hamiltonian to a Hermitian
one by introducing a similarity transformation on the particle operators.
The symmetry of the system leads to the pairing coalescence of the energy
levels in the unidirectional limit. In subsection \ref{IIb}, we investigate
the dynamics at EP and then high-order EP. In section \ref{III}, we focus on
the applications of our finding on a spin-$1/2$ fermion system. We simulate
the dynamic magnetization on a square lattice. In section \ref{IV}, we
summarize the results.

\begin{figure}[tbh]
\centering \includegraphics[width=0.45\textwidth]{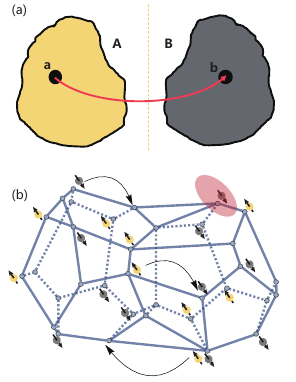}
\caption{(a) Schematic illustration of the structure of tight-binding model
concerned in this work, which supports exceptional spectrum. It consists of
two identical sub-systems (two shadow areas A and B, yellow and gray,
respectively), which have a reflectional symmetry about the axis (dotted
line). Such a system has degeneracy spectrum in the absence of
particle-particle interaction. The non-Hermiticity arises from a
unidirectional hopping between two symmetric points \textrm{a} and \textrm{b}
, which is indicated by the red arrow. It is shown that the non-Hermitian
term results in a Jordan block in each subspace spanned by two degenerate
states. (b) Schematic of an example system, which is a non-interacting
itinerant electron system, described by the Hamiltonian in equation (\protect
\ref{H spin}). The sub-Hamiltonian for electron with spin up (yellow)
corresponds to yellow sub-system in (a), while the spin down (gray)
corresponds to the gray part.\ The non-Hermitian term [red arrow in (a)] is
realized by a local critical complex field shaded red.}
\label{fig1}
\end{figure}

\section{General formalism}

\label{II}

In this section, we present a general formalism\ of systems supporting the
coalescing spectrum. Firstly, we construct a class of non-Hermitian
Hamiltonians which possesses pseudo-Hermiticity and inversion symmetry. Pseudo-Hermitian Hamiltonian has a real spectrum or else its
complex eigenvalues always occur in complex conjugate pairs \cite{MA}. A
pseudo-Hermitian Hamiltonian satisfies the condition%
\begin{equation}
SHS^{-1}=H^{\dagger },
\end{equation}%
where $S$ is a Hermitian linear automorphism \cite{SSP}. Secondly, we show
that such Hamiltonians can be tuned to have full coalescing spectrum and
then support a special dynamics.

\subsection{Model and solution}

\label{IIa}

We start with a class of non-Hermitian Hamiltonians in the form

\begin{eqnarray}
H &=&H_{1}+H_{2}+H_{12},  \label{H} \\
H_{\lambda } &=&\sum_{i\geqslant j}^{N}J_{ij}a_{i,\lambda }^{\dag
}a_{j,\lambda }+\mathrm{H.c.},(\lambda =1,2), \\
H_{12} &=&\sum_{j=1}^{N}\kappa _{j}a_{j,1}^{\dag }a_{j,2}+\gamma
^{2}\sum_{j=1}^{N}\kappa _{j}a_{j,2}^{\dag }a_{j,1},
\end{eqnarray}%
which consists of two identical Hermitian clusters $H_{\lambda }$ with $%
\lambda =1,2$, respectively. The structure of the model is schematically
illustrated in figure \ref{fig1}(a). The non-Hermiticity arises from the
real asymmetrical hopping strength with $\gamma \neq 1$. The
distribution of the hopping integrals $\left\{ \kappa _{j}\right\} $
determines the non-Hermitian term to be a macroscopic or a local term. Here 
$a_{j,\lambda }^{\dag }$\ ($a_{j,\lambda }$) is the boson or fermion
creation (annihilation) operator at the $i$th site in the $\lambda $th
cluster.
In this work, we only consider unidirectional hopping with the same position for the following reasons: (i) This allows us to perform analytical analysis. (ii)
Such a term corresponds to the description of magnetic impurity in the
following section. (iii) When the interaction between different positions is
considered, the exceptional spectrum may also exists or not, depending on
the structure of $H_{\lambda }$, according to the theorem in
reference {\cite{WPPRB}}. The cluster $H_{\lambda }$ is defined by the
distribution of the hopping integrals $\left\{ J_{ij}\right\} $ with $i>j$\
and on-site potentials $\left\{ J_{jj}\right\} $. The matrix $J_{ij}$ is
Hermitian and we only consider the case with $\gamma >0$\ in this paper. Two
Hamiltonians $H_{\lambda }$ have the same eigenfunctions and real spectral
structures and the whole Hamiltonian is not self-adjoint except the case
with $\gamma =1$. In the following, we will show that it still has full real
spectrum and $\left( N+1\right) $-order exceptional point at $\gamma =0$
(see \ref{AppendixB}), near which the dynamics of the Hamiltonian $H$\ is
crucial to the conclusion of this paper.

At first, we consider a case with $\gamma =0$ and $\kappa _{j}=\kappa $,
representing uniform unidirectional hopping between two identical clusters.
The Hermitian Hamiltonians $H_{\lambda }$\ can be written in the diagonal
form%
\begin{equation}
H_{\lambda }=\sum_{k=1}^{N}\varepsilon _{k}A_{k,\lambda }^{\dag
}A_{k,\lambda }
\end{equation}%
by the transformation%
\begin{equation}
A_{k,\lambda }=\sum_{j=1}^{N}g_{k,j}a_{j,\lambda },
\end{equation}%
where we use $k$ to denote the index of eigenmode of $H_{\lambda }$, which
becomes wave vector for a system with translational symmetry. Here the
spectrum $\varepsilon _{k}$\ is real and $g_{k,j}$\ satisfies orthonormal
relations%
\begin{equation}
\sum_{j=1}^{N}\left( g_{k,j}\right) ^{\ast }g_{k^{\prime },j}=\delta
_{kk^{\prime }},\sum_{k=1}^{N}\left( g_{k,i}\right) ^{\ast }g_{k,j}=\delta
_{ij},
\end{equation}%
and can be obtained from the diagonalization of the $N\times N$ matrix $%
\left\{ J_{ij}\right\} $. Accordingly we have%
\begin{equation}
H_{12}=\kappa \sum_{j=1}^{N}a_{j,1}^{\dag }a_{j,2}=\kappa
\sum_{k=1}^{N}A_{k,1}^{\dag }A_{k,2}
\end{equation}%
which allows the block diagonal form of the Hamiltonian%
\begin{eqnarray}
H &=&\sum_{k=1}^{N}H_{k} \\
H_{k} &=&\varepsilon _{k}\left( A_{k,1}^{\dag }A_{k,1}+A_{k,2}^{\dag
}A_{k,2}\right) +\kappa A_{k,1}^{\dag }A_{k,2} \\
&=&\left( 
\begin{array}{ll}
A_{k,1}^{\dag } & A_{k,2}^{\dag }%
\end{array}%
\right) h_{k}\left( 
\begin{array}{l}
A_{k,1} \\ 
A_{k,2}%
\end{array}%
\right) .  \label{App}
\end{eqnarray}%
due to the relation $\left[ H_{k},H_{k^{\prime }}\right] =0$. Importantly,
matrix

\begin{equation}
h_{k}=\left( 
\begin{array}{ll}
\varepsilon _{k} & \kappa \\ 
0 & \varepsilon _{k}%
\end{array}
\right)
\end{equation}%
is a Jordan block. The coalescing vector is $\left( 
\begin{array}{ll}
1 & 0%
\end{array}%
\right)^T$, while the auxiliary vector is $\left( 
\begin{array}{ll}
0 & 1%
\end{array}%
\right)^T$. Then the eigenstate set of $H$\ is identical to that of $H_{1}$,
while the auxiliary set is the eigenstate set of $H_{2}$.

Secondly, in the case with nonzero $\gamma $ and non-uniform $\kappa _{j}$,
the Hamiltonian can be written as the following form when $H_{\lambda }$\ is
a bipartite lattice (see \ref{AppendixA}),

\begin{equation}
H=\sum_{n=1,\rho =\pm }^{N}E_{\rho ,n}\overline{d}_{n,\rho }d_{n,\rho }.
\end{equation}%
We note that when taking $\gamma \rightarrow 0$\ we have $\overline{d}%
_{n,\pm }\sim \sum_{j=1}^{N}f_{n,\pm }^{j}a_{j,1}^{\dag }$ and $d_{n,\pm
}\sim \sum_{j=1}^{N}\left( f_{n,\pm }^{j}\right) ^{\ast }a_{j,2}$, with $%
f_{n,+}^{j}\longrightarrow f_{n,-}^{j}$, i.e., the pair of operators with
subscript $\pm $\ coalesce, which accord with the analysis in the case of $%
\gamma =0$ and $\kappa _{j}=\kappa $. The most fascinating feature of such
systems is that even a single unidirectional tunneling (the simplest
configuration for non-uniform $\kappa _{j}$) can result in exceptional
spectrum.

\subsection{Exceptional spectrum and critical dynamics}

\label{IIb}

Now we turn to the dynamics driven by the non-Hermitian Hamiltonian. At
first, we still consider a case with $\gamma =0$ and $\kappa _{j}=\kappa $,
in which the Hamiltonians $H$\ has been written as the sum of $N$
independent sub-Hamiltonians $H_{k}$. In the case of zero $\kappa $, $H$\
has two-fold degeneracy spectrum, which is similar to the Kramer's
degeneracy. However, when $\kappa $\ is switched on, each pairs of
degenerate energy levels coalesce into a single one. The original Kramer's
spectrum transits to an exceptional spectrum. As known that a system at EP
exhibits unusual dynamics. It is presumable that a system with exceptional
spectrum should supports interesting dynamical phenomena, particularly in
many-fermion system.

The dynamics for any initial state is governed by the time evolution
operator $U(t)=e^{-iHt}$. It is expressed explicitly as%
\begin{eqnarray}
U(t) &=&\prod_{k}U_{k}(t)=\prod_{k}e^{-iH_{k}t} \\
&=&e^{-i\sum_{k}\varepsilon _{k}\left( A_{k,1}^{\dag }A_{k,1}+A_{k,2}^{\dag
}A_{k,2}\right) t}\prod_{k}e^{-i\kappa A_{k,1}^{\dag }A_{k,2}t}.
\end{eqnarray}%
For fermion system, we have it reduces to%
\begin{equation}
U(t)=\prod_{k}(1-i\kappa A_{k,1}^{\dag }A_{k,2}t),
\end{equation}%
where an overall dynamical phase factor is omitted and the identity $%
(A_{k,1}^{\dag }A_{k,2})^{2}=0$\ for fermion is used. Then for a given
initial state $\left\vert \Phi (0)\right\rangle =A_{k,2}^{\dag }\left\vert 
\mathrm{Vac}\right\rangle $, we have%
\begin{equation}
\left\vert \Phi (t)\right\rangle =\left( A_{k,2}^{\dag }-i\kappa
A_{k,1}^{\dag }t\right) \left\vert \mathrm{Vac}\right\rangle
\end{equation}%
which indicates $\left\vert \Phi (\infty )\right\rangle \rightarrow
A_{k,1}^{\dag }\left\vert \mathrm{Vac}\right\rangle $, i.e., transferring an
eigenstate of $H_{2}$\ to that of $H_{1}$. It can be extended to an
arbitrary initial state. We are interested in a typical many-particle
initial state%
\begin{equation}
\left\vert \Phi (0)\right\rangle =\prod_{j=1}^{N}a_{j,2}^{\dag }\left\vert 
\mathrm{Vac}\right\rangle =\prod_{k}^{N}A_{k,2}^{\dag }\left\vert \mathrm{Vac%
}\right\rangle ,
\end{equation}%
which is fully occupied state of system $H_{2}$. In the many-fermion system,
high-order exceptional point occurs(see \ref{AppendixB}). Similarly, at
large $t$ limit, we have%
\begin{equation}
\left\vert \Phi (t)\right\rangle \propto \prod_{k}^{N}A_{k,1}^{\dag
}\left\vert \mathrm{Vac}\right\rangle =\prod_{j=1}^{N}a_{j,1}^{\dag
}\left\vert \mathrm{Vac}\right\rangle ,
\end{equation}%
i.e., a fully occupied state of system $H_{1}$. We can
calculate the number of the fermions in lattice $1$ and $2$%
\begin{equation}
n_{\lambda }=\sum_{j=1}^{N}\frac{\left\langle \Phi (t)\right\vert
a_{j,\lambda }^{\dagger }a_{j,\lambda }\left\vert \Phi (t)\right\rangle }{%
\left\vert \left\langle \Phi (t)|\Phi (t)\right\rangle \right\vert }
,(\lambda =1,2),
\end{equation}%
respectively. It indicates that all the fermions in lattice $2$ transfer to
lattice $1$, eventually. We find that during the dynamical process, (i) the
total particle number is always conservative, (ii) the particle number in
lattice $2$ vanishes after sufficient long time for an arbitrary initial
state, (iii) it tends to the final state with speed in power law, and the
exponent\ equals to the initial particle number in lattice $2$, the order of
EP. Based on these analyses, we conclude that no matter what types of
initial state, pure state or mixed state, the final state is a state with a
fixed particle number in lattice $1$, which is determined by the maximal
particle number in lattice $2$\ among all the components of the initial
state. For the case with non-uniform $\kappa _{j}$, the solutions we
obtained benefit to the numerical simulation for many-fermion system,
avoiding the exponentially increasing computer time as a function of the
lattice size. We will see that the main result is independent of the
distribution of $\left\{ \kappa _{j}\right\} $.

\begin{figure*}[tbh]
\centering
\includegraphics[width=0.98\textwidth]{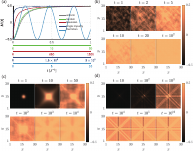}
\caption{Numerical results for the dynamic magnetization process of a $%
30\times 30$ square lattice with uniform \textrm{NN} hopping $J_{ij}=J=1$.
(a) The average magnetizations $M(t)$ as functions of time for different
non-Hermitian field distributions. As a comparison, the result of $M(t)$ for
a uniform Hermitian field $\mathbf{B}_{j}=\protect\kappa_{0} (1,0,0)$ is
also presented. Panels (b)-(d) are snapshots of the magnetizations $m(x,y,t)$
with three representative non-Hermitian field distributions: random,
Gaussian and single impurity, respectively. Other parameters are $\protect%
\kappa _{0}=1$, $\protect\alpha =0.16$ and $(x_{0},y_{0})=(17,16)$.}
\label{fig2}
\end{figure*}

\section{Dynamic magnetization}

\label{III}

In this section we apply the obtained result to a spin-$1/2$ fermionic model
with complex impurities [see figure \ref{fig1}(b)]. The Hamiltonian on an
arbitrary lattice can be written as

\begin{equation}
H_{\mathrm{F}}=\sum_{i\geqslant j}^{N}\sum_{\sigma =\uparrow ,\downarrow
}J_{ij}c_{i,\sigma }^{\dag }c_{j,\sigma }+\mathrm{H.c.}+\sum_{j=1}^{N}%
\mathbf{B }_{j}\cdot \mathbf{s}_{j},  \label{H spin}
\end{equation}%
where operator $c_{i,\sigma }^{\dag }$\ creates a fermion of spin $\sigma $
at site $i$, and $\mathbf{s}_{j}=(s_{j}^{x},s_{j}^{y},s_{j}^{z})$\ is the
spin-$1/2$ operator, which is defined by%
\begin{eqnarray}
s_{j}^{+} &=&\frac{1}{2}\left( s_{j}^{x}+is_{j}^{y}\right) =\left(
s_{j}^{-}\right) ^{\dag }=c_{j,\uparrow }^{\dag }c_{j,\downarrow } \\
s_{j}^{z} &=&\frac{1}{2}\left( c_{j,\uparrow }^{\dag }c_{j,\uparrow
}-c_{j,\downarrow }^{\dag }c_{j,\downarrow }\right)
\end{eqnarray}%
satisfying the Lie algebra commutation relations,

\begin{equation}
\left[ s^{+},s^{-}\right] =2s^{z},\left[ s^{z},s^{\pm }\right] =\pm s^{\pm }.
\end{equation}%
Here $J_{ij}$\ for $i\neq j$\ is hopping strength between two sites ($i,j$),
and $J_{jj}$ is on-site potential. $\mathbf{B}_{j}$\ is on-site complex
magnetic field, inducing non-Hermitian impurities. The motivation of
introducing a complex magnetic field is to simulating the classical
magnetization process in the framework of quantum mechanics. As well known,
the friction must be involved in the theory of magnetization based on the
classical physics. Non-Hermitian Hamiltonian with a complex magnetic field
may be a good candidate for this task.

In history, a complex field was usually induced to characterize
the connections between a closed system to the environment
phenomenologically. Recently, the topic of complex field has been
investigated from many aspects.Theoretically, a non-Hermitian Hamiltonian is
the reduced description for a selected sub-system of a Hermitian system,
where the complementary subspace is take into account by means of an
effective interaction described by a non-Hermitian complex potential \cite%
{Muga,Jin2,Jin}. Experimentally, a complex field is usually simulated in the
optical system with gain and loss \cite{Gupta,Guo,Makris}. In addition,
there is an experimental protocol for complex magnetic field via atomic
system, referred to as "heralded Magnetism"\cite{LEE}. Very recently, the
non-Hermiticity in real quantum systems has been experimentally demonstrated 
\cite{REN,Liu,WU,Partanen,Li}.

In this work, we only consider the field in the form 
\begin{equation}
\mathbf{B}_{j}=\kappa _{j}(1,i,0),
\end{equation}%
with several typical distribution of $\left\{ \kappa _{j}\right\} $. $%
B_{j}\cdot s_{j}$ is the non-Hermitian term. It can drive the fermion from
spin-down state $c_{j,\downarrow }^{\dag }\left\vert \mathrm{Vac}%
\right\rangle $ to the spin-up state $c_{j,\uparrow }^{\dag }\left\vert 
\mathrm{Vac}\right\rangle $ at $j$-th site. The crucial point is that the
local system\ $\mathbf{B}_{j}\cdot \mathbf{s}_{j}$\ is at EP. Remarkably,
the result in above section tells us that it may induce multi-EP of the
whole system. The fermionic Hamiltonian $H_{\mathrm{F}}$ can be expressed in
the form of $H$ in equation (\ref{H}) by taking\ $c_{j,\uparrow }=a_{j,1}$\
and $c_{j,\downarrow }=a_{j,2}$, respectively. Accordingly, the inversion
operator $\mathcal{P}$\ (see \ref{AppendixA}) maps to spin-reversal operator 
$\mathcal{S}$, where $\mathcal{S}$ has the action $\mathcal{S}c_{j,\sigma
}^{\dag }$($c_{j,\sigma }$)$\mathcal{S}^{-1}=c_{j,-\sigma }^{\dag }$($%
c_{j,-\sigma }$). A straightforward result can be obtained from the analysis
in last section.

We are interested in the dynamic process of magnetization driven by
non-Hermitian magnetic field. We first consider the simplest case with
uniform field distribution $\mathbf{B}_{j}=\kappa (1,i,0)$. The system we
concern is half-filled and the initial state is a saturated ferromagnetic
state%
\begin{equation}
\left\vert \Phi (0)\right\rangle =\left\vert \Downarrow \right\rangle
=\prod_{j=1}^{N}c_{j,\downarrow }^{\dag }\left\vert \mathrm{Vac}%
\right\rangle ,
\end{equation}%
which is a popular state in physics. Here we choose this initial state for
the convenience of calculation. The evolved state driven by the Hamiltonian
with non-Hermitian magnetic field is%
\begin{equation}
\left\vert \Phi (t)\right\rangle =\prod_{k=1}^{N}(1-i\kappa C_{k,\uparrow
}^{\dag }C_{k,\downarrow }t)\prod_{j=1}^{N}c_{j,\downarrow }^{\dag
}\left\vert \mathrm{Vac}\right\rangle ,
\end{equation}%
which shows that the amplitude of the evolved state is dominant by the term
with $t^{N}$ for a long time. Here an overall dynamical phase factor is
omitted and%
\begin{equation}
C_{k,\sigma }=\sum_{j=1}^{N}g_{k,j}c_{j,\sigma }.
\end{equation}%
We employ magnetization%
\begin{equation}
m_{j}(t)=\frac{\left\langle \Phi (t)\right\vert s_{j}^{z}\left\vert \Phi
(t)\right\rangle }{\left\vert \left\vert \Phi (t)\right\rangle \right\vert
^{2}},
\end{equation}%
and average magnetization%
\begin{equation}
M(t)=\frac{1}{N}\sum_{j=1}^{N}m_{j}(t),
\end{equation}%
to characterize the dynamic magnetization process. Based on the identity%
\begin{equation}
\prod_{j=1}^{N}c_{j,\downarrow }^{\dag }\left\vert \mathrm{Vac}\right\rangle
=\prod_{k=1}^{N}C_{k,\downarrow }^{\dag }\left\vert \mathrm{Vac}%
\right\rangle ,
\end{equation}%
which are two expressions of a same state with fully filled
spin-up electrons, in real space and collective mode space, respectively. a
straightforward derivation results in

\begin{equation}
M\left( t\right) =\frac{\sum_{n=0}^{N}(2n-N)C_{N}^{n}(\kappa t)^{2n}}{\left(
2N\right) \sum_{n=0}^{N}C_{N}^{n}(\kappa t)^{2n}}=\frac{(\kappa t)^{2}-1}{2 %
\left[ (\kappa t)^{2}+1\right] },  \label{MT}
\end{equation}

$M\left( t\right) $ is plotted in figure \ref{fig2}(a), in comparison with
numerical results for various situations. It is clear that after a long
time, $\left\vert \Phi (t)\right\rangle $ tends to another saturated
ferromagnetic state 
\begin{equation}
\left\vert \Uparrow \right\rangle =\prod_{k=1}^{N}C_{k,\uparrow }^{\dag
}\left\vert \mathrm{Vac}\right\rangle =\prod_{j=1}^{N}c_{j,\uparrow }^{\dag
}\left\vert \mathrm{Vac}\right\rangle ,  \label{state_up}
\end{equation}%
with $M\left( \infty \right) =1/2$. In fact, no matter what the initial
state is, after a long time, the evolved state tends to the state in
equation (\ref{state_up}) due to the EP dynamics.

\begin{figure*}[tbh]
\centering
\includegraphics[width=0.98\textwidth]{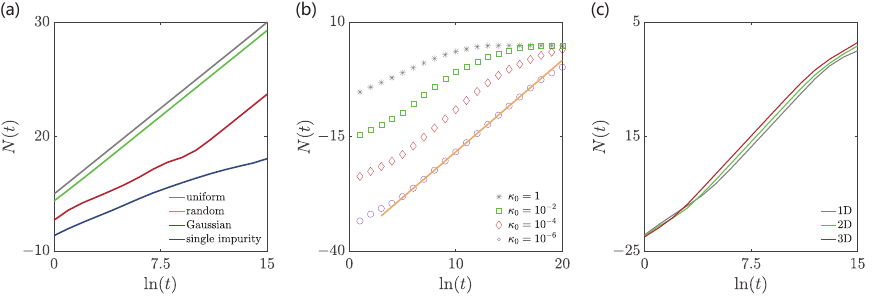}
\caption{Plots of $\mathrm{N}(t)$ defined in equation (\protect\ref{Nt}) as
functions of $\ln (t)$. (a) $\mathrm{N}(t)$ for a $30\times 30$ square
lattice with different non-Hermitian field distributions, which correspond
to the processes plotted in figure \protect\ref{fig2}(a). (b) $\mathrm{N}(t)$
for a $20\times 20$ square lattice with single impurity $\protect\kappa _{0} 
\protect\delta (x-11)\protect\delta (y-10)$ under different field strength $%
\protect\kappa _{0}$. The orange line is $\mathrm{N}(t)=2\ln t-38.2$, which
indicates that for a smaller $\protect\kappa _{0}$, the speed of
magnetization is closer to that of uniform field distribution. (c) $\mathrm{%
N }(t)$ for the lattices of different dimension (1D, 2D and 3D) with single
impurity of field strength $\protect\kappa _{0}=1$. The size of three
lattices are $64$, $8\times 8$, and $4\times 4\times 4$, The positions of
the impurity of the 2D and 3D lattices are $(4,1)$ and $(2,3,1)$,
respectively.}
\label{fig3}
\end{figure*}

It is interesting to investigate what happens when $\left\{ \kappa
_{j}\right\} $ is taken as non-uniform distributions in two-dimensional(2D)
square lattice with uniform nearest neighbor (NN) $\left\langle
i,j\right\rangle $\ hopping strength $J_{ij}=1$. Numerical simulations are
performed for three typical forms: (i) Random distribution $\kappa
(x,y)=\kappa _{0}\mathrm{Ran}[0,1]$, where \textrm{Ran}$[0,1]$ denotes
uniform random real number within the interval $[0,1]$, (ii) Gaussian
distribution $\kappa (x,y)=\kappa _{0}\exp \{-\alpha
^{2}[(x-x_{0})^{2}+(y-y_{0})^{2}]\}$, and (iii) Single impurity $\kappa
(x,y)=\kappa _{0}\delta (x-x_{0})\delta (y-y_{0})$. Numerical simulations
for the dynamical processes of magnetization\ are performed for such three
representative cases. Numerical results for $M(t)$ and $m(x,y,t)=m_{j}(t)$\
[the coordinate of position $j$ in 2D square lattice is $(x,y)$]\ are
plotted in figure \ref{fig2}(a) and figure \ref{fig2}(b)-\ref{fig2}(d),
respectively. We find that the itinerant electron system with different
distributions of non-Hermitian magnetic field have similar dynamical
processes and all tend to the same saturated ferromagnetic state $\left\vert
\Uparrow \right\rangle =\prod_{j=1}^{N}c_{j,\uparrow }^{\dag }\left\vert 
\mathrm{Vac}\right\rangle $ with $M\left( \infty \right) =1/2$ finally. In
contrast, for the case of uniform Hermitian field, $M(t)$ is a periodic
function. To measure the speed of the magnetization\ we introduce a
dimensionless quantity $\mathrm{N}\left( t\right) $

\begin{equation}
\mathrm{N}\left( t\right) =\ln \frac{1+2M\left( t\right) }{1-2M\left(
t\right) }.  \label{Nt}
\end{equation}%
From equation (\ref{MT}), we find that for the case with uniform field
distribution we simply have\textbf{\ }$\mathrm{N}\left( t\right) =2\ln
\kappa +2\ln t$\textbf{, }where factor $2$\ characterize the speed. $\mathrm{%
\ N}(t)$ for the processes in figure \ref{fig2} are plotted in figure \ref%
{fig3}(a). It shows that the final results are independent of the
distributions of $\kappa (x,y)$, which only affects the speed of the
magnetization. Figure \ref{fig3}(b) indicates that for a smaller $\kappa
_{0} $, the speed of magnetization is closer to that of uniform field
distribution. We also investigate the effect of dimensionality on the speed
of the magnetization. We compute $\mathrm{N}(t)$ for several typical cases,
including $1$D periodic chain, $2$D square lattice and $3$D cubic lattice
with single non-Hermitian impurity [see figure \ref{fig3}(c)]. It indicates
that the result is independent of the dimensionality approximately.

\section{Summary}

\label{IV}

In summary, we have developed a theory for a class of non-Hermitian
Hamiltonian which supports a special dynamics due to the appearance of
exceptional full real spectrum. The most fascinating feature of such systems
is that even a single unidirectional tunneling can result in exceptional
spectrum. It is shown that a macroscopic effect can be induced by a local
interaction in a non-Hermitian system. To demonstrate this point, we apply
the theory on an itinerant electron system subjected to complex fields. As
an application, we have studied the dynamic magnetization induced by the
field. It shows that an initial saturated ferromagnetic state at
half-filling can be driven into its opposite state according to the dynamics
of high-order exceptional point. Numerical simulations for several
representative situations, including lattice dimensions, global random and
local impurity distributions indicate that the dynamic magnetization
processes exhibit universal behavior. Our findings provide alternative
explanation for dynamic magnetization process in itinerant electron systems
in the context of non-Hermitian quantum mechanics.

\ack This work was supported by the National Natural Science
Foundation of China (under Grant No. 11874225).

\appendix

\section{The derivations of the Hamiltonian and the coalescing spectrum}

\label{AppendixA} \setcounter{equation}{0} \renewcommand{\theequation}{A	%
\arabic{equation}} \renewcommand{\thesubsection}{\arabic{subsection}}

In this Appendix, we present a derivation on the solution of the Hamiltonian 
$H$ in equation (\ref{H}) on a bipartite lattice with nonzero $\gamma $ and
show that the coalescing spectrum appears as $\gamma $\ truns to zero.

To this end, we introduce a set of particle operators $\overline{d}%
_{j,\lambda }$ and $d_{j,\lambda }$ \cite{ZXZ1}.

\begin{eqnarray}
\overline{d}_{j,1} &=&\frac{1}{\sqrt{\gamma }}a_{j,1}^{\dag },d_{j,1}=\sqrt{%
\gamma }a_{j,1}, \\
\overline{d}_{j,2} &=&\sqrt{\gamma }a_{j,2}^{\dag },d_{j,2}=\frac{1}{\sqrt{%
\gamma }}a_{j,2},  \label{transformation}
\end{eqnarray}%
which are canonical conjugate pairs, satisfying%
\begin{eqnarray}
\lbrack d_{j,\lambda },\overline{d}_{i,\lambda ^{\prime }}]_{\pm } &=&\delta
_{ij}\delta _{\lambda \lambda ^{\prime }}, \\
\lbrack d_{j,\lambda },d_{i,\lambda ^{\prime }}]_{\pm } &=&[\overline{d}%
_{j,\lambda },\overline{d}_{i,\lambda ^{\prime }}]_{\pm }=0,
\end{eqnarray}%
where $[\cdot ,\cdot ]_{\pm }$ denotes the commutator and anti-commutator.
Transformation in equation (\ref{transformation}) is essentially a
similarity transformation with a singularity at $\gamma =0$, beyond which\
it allows us to rewrite the Hamiltonians in the form%
\begin{eqnarray}
H_{\lambda } &=&\sum_{i\geqslant j}^{N}J_{ij}\overline{d}_{i,\lambda
}d_{j,\lambda }+\sum_{i\geqslant j}^{N}\left( J_{ij}\right) ^{\ast }%
\overline{d}_{j,\lambda }d_{i,\lambda } \\
H_{12} &=&\gamma \sum_{j=1}^{N}\kappa _{j}\overline{d}_{i,1}d_{j,2}+\gamma
\sum_{j=1}^{N}\left( \kappa _{j}\right) ^{\ast }\overline{d}_{j,2}d_{i,1}.
\end{eqnarray}%
More explicitly we have%
\begin{equation}
H=\sum_{i,j=1}^{N}\sum_{\lambda ,\lambda ^{\prime }=1}^{2}\overline{d}%
_{i,\lambda }\mathcal{H}_{i,j,\lambda ,\lambda ^{\prime }}d_{j,\lambda
^{\prime }},
\end{equation}%
where the core matrix is%
\begin{eqnarray}
\mathcal{H} &=&\sum_{i\geqslant j}^{N}\sum_{\lambda =1,2}J_{ij}\left\vert
i,\lambda \right\rangle \overline{\left\langle j,\lambda \right\vert }%
+\sum_{i\geqslant j}^{N}\sum_{\lambda =1,2}\left( J_{ij}\right) ^{\ast
}\left\vert j,\lambda \right\rangle \overline{\left\langle i,\lambda
\right\vert } \\
&&+\gamma \sum_{j=1}^{N}\kappa _{j}\left\vert j,1\right\rangle \overline{%
\left\langle j,2\right\vert }+\gamma \sum_{j=1}^{N}\left( \kappa _{j}\right)
^{\ast }\left\vert j,2\right\rangle \overline{\left\langle j,1\right\vert }.
\end{eqnarray}%
based on the orthonormal complete basis $\left\{ \left\vert i,\lambda
\right\rangle \right\} =\left\{ \overline{d}_{i,\lambda }\left\vert \mathrm{%
\ Vac}\right\rangle \right\} $and $\left\{ \overline{\left\langle i,\lambda
\right\vert }\right\} =\left\{ \left\langle \mathrm{Vac}\right\vert
d_{i,\lambda }\right\} $($\left\vert \mathrm{Vac}\right\rangle $ is the
vacuum state of operator $a_{j,\lambda }$), satisfying%
\begin{equation}
\overline{\left\langle i,\lambda \right\vert }j,\lambda ^{\prime }\rangle
=\delta _{ij}\delta _{\lambda \lambda ^{\prime }}.
\end{equation}%
We find that the matrix $\mathcal{H}$\ is Hermitian in the basis set $%
\left\{ \left\vert i,\lambda \right\rangle ,\overline{\left\langle i,\lambda
\right\vert }\right\} $, although $\left( \left\vert i,\lambda \right\rangle
\right) ^{\dag }\neq \overline{\left\langle i,\lambda \right\vert }$. For the sake of simplicity, we only need to diagonalize such a
Hermitian matrix 
\begin{equation}
h=\sum_{i\geqslant j}^{N}\sum_{\lambda =1,2}J_{ij}\left\vert i,\lambda
\right\rangle \left\langle j,\lambda \right\vert +\gamma
\sum_{j=1}^{N}\kappa _{j}\left\vert j,1\right\rangle \left\langle
j,2\right\vert +\mathrm{H.c.},
\end{equation}%
with $\left\langle i,\lambda \right\vert j,\lambda ^{\prime }\rangle =\delta
_{ij}\delta _{\lambda \lambda ^{\prime }}.$ Importantly, $h$\ has inversion
symmetry due to the reality of matrix elements $\left\{ \kappa _{j}\right\} $%
, i.e.,%
\begin{equation}
\mathcal{P}h\mathcal{P}^{-1}=h,
\end{equation}%
where $\mathcal{P}$ has the action $\mathcal{P}\left\vert j,1\right\rangle
=\left\vert j,2\right\rangle $ ($\mathcal{P}\left\vert j,2\right\rangle
=\left\vert j,1\right\rangle $). Note that such symmetry is independent of
the structure of the sub-lattice determined by $\left\{ J_{ij}\right\} $.
Then we can rewrite the matrix in the block diagonal form%
\begin{equation}
h=h_{+}+h_{-},
\end{equation}%
where%
\begin{equation}
h_{\pm }=\sum_{i\geqslant j}^{N}J_{ij}\left\vert i,\pm \right\rangle
\left\langle j,\pm \right\vert +\mathrm{H.c.}\pm \gamma \sum_{j=1}^{N}\kappa
_{j}\left\vert j,\pm \right\rangle \left\langle j,\pm \right\vert ,
\end{equation}%
and%
\begin{equation}
\left\vert j,\pm \right\rangle =\frac{\left\vert j,1\right\rangle \pm
\left\vert j,2\right\rangle }{\sqrt{2}}.
\end{equation}%
We note that $h_{+}$ and $h_{-}$ satisfy $\left[ h_{+},h_{-}\right] =0$\ and
the matrice of $h_{+}$ and $h_{-}$ become identical as $\gamma \rightarrow 0$%
. Accordingly, the solution of the Schrodinger equation

\begin{equation}
h_{\pm }\left\vert \psi _{\pm ,n}\right\rangle =E_{\pm ,n}\left\vert \psi
_{\pm ,n}\right\rangle ,
\end{equation}%
has the form%
\begin{equation}
\left\vert \psi _{\pm ,n}\right\rangle =\sum_{j=1}^{N}f_{n,\pm }^{j}\frac{%
\left\vert j,1\right\rangle \pm \left\vert j,2\right\rangle }{\sqrt{2}},
\end{equation}%
satisfying%
\begin{equation}
\mathcal{P}\left\vert \psi _{\pm ,n}\right\rangle =\pm \left\vert \psi _{\pm
,n}\right\rangle .
\end{equation}%
Then the Hamiltonian is diagonalized as the form

\begin{equation}
H=\sum_{n=1,\rho =\pm }^{N}E_{\rho ,n}\overline{d}_{n,\rho }d_{n,\rho },
\end{equation}%
where%
\begin{eqnarray}
\overline{d}_{n,\pm } &=&\sum_{j=1}^{N}f_{n,\pm }^{j}\left( \frac{1}{\sqrt{
\gamma }}a_{j,1}^{\dag }\pm \sqrt{\gamma }a_{j,2}^{\dag }\right) , \\
d_{n,\pm } &=&\sum_{j=1}^{N}\left( f_{n,\pm }^{j}\right) ^{\ast }\left( 
\sqrt{\gamma }a_{j,1}\pm \frac{1}{\sqrt{\gamma }}a_{j,2}\right) .
\end{eqnarray}%
Although the above solution is only true for nonzero $\gamma $, one can
extrapolate the approximate solution at $\gamma =0$\ by taking $\gamma
\rightarrow 0$. In the limit of zero $\gamma $, we have $f_{n,+}^{j}%
\longrightarrow f_{n,-}^{j}\rightarrow g_{k,j}$, which results in

\begin{equation}
\overline{d}_{n,\pm }\sim \sum_{j=1}^{N}g_{k,j}a_{j,1}^{\dag },d_{n,\pm
}\sim \sum_{j=1}^{N}\left( g_{k,j}\right) ^{\ast }a_{j,2}.
\end{equation}

We demonstrate the result by a simple system, which consists of two
identical $N$-site uniform chain with a unidirectional hopping between them.
The Hamiltonian has the form 
\begin{eqnarray}
H &=&H_{1}+H_{2}+H_{12}, \\
H_{\lambda } &=&\sum_{i=1}^{N-1}Ja_{i,\lambda }^{\dagger }a_{i+1,\lambda }+ 
\mathrm{H.c.},(\lambda =1,2) \\
H_{12} &=&\kappa a_{1,1}^{\dagger }a_{1,2}.
\end{eqnarray}
According to our analysis above, the eigenvalues and eigenvectors in
single-particle invariant subspace have the form{\ 
\begin{eqnarray}
\epsilon _{1,n} &=&2J\cos \left( \frac{n\pi }{N+1}\right) , \\
\left\vert \psi _{1,n}\right\rangle &=&\sqrt{\frac{2}{N+1}}
\sum_{j=1}^{N}\sin \left( \frac{n\pi }{N+1}j\right) a_{j,1}^{\dagger
}\left\vert \mathrm{Vac}\right\rangle ,
\end{eqnarray}
which satisfies 
\begin{equation}
{H}\left\vert \psi _{1,n}\right\rangle =\epsilon _{1,n}\left\vert \psi
_{1,n}\right\rangle .
\end{equation}
{Similarly, for the system }${H}^{\dag }${, we have 
\begin{equation}
\left\vert \varphi _{1,n}\right\rangle =\sqrt{\frac{2}{N+1}}
\sum_{j=1}^{N}\sin \left( \frac{n\pi }{N+1}j\right) a_{j,2}^{\dagger
}\left\vert \mathrm{Vac}\right\rangle ,
\end{equation}
which satisfies } 
\begin{equation}
{H}^{\dag }\left\vert \varphi _{1,n}\right\rangle =\epsilon _{1,n}\left\vert
\varphi _{1,n}\right\rangle .
\end{equation}
Obviously we have 
\begin{equation}
\langle \varphi _{1,m}\left\vert \psi _{1,n}\right\rangle =0,
\end{equation}
for any $m$ and $n$, which indicates that $\left\vert \psi
_{1,n}\right\rangle $\ is coalscing states due to the vanishing biorthogonal
norm \cite{NM}. Furthermore, }this conclusion can be true for the case with
multiple\ unidirectional hoppings.

\section{High-order exceptional point}

\label{AppendixB} \setcounter{equation}{0} \renewcommand{\theequation}{B	%
\arabic{equation}} \renewcommand{\thesubsection}{\arabic{subsection}}

In this Appendix, we prove that our system with more than one fermion has
high-order exceptional point.

We consider the following case. In the $N$-fermion invariant subspace with
each $k$ filled by one fermion. In this $N$-fermion basis, the matrix
representation of Hamiltonian $H$\ in equation (\ref{App}) can be written as%
\textbf{\ }%
\begin{equation}
h=T_{N}+\sum_{k=1}^{N}\varepsilon _{k}.
\end{equation}%
Here $T_{N}$\ can be expressed in the form%
\begin{eqnarray}
T_{N} &=&I_{2}\otimes T_{N-1}+T_{1}\otimes I_{2^{N-1}} \\
&=&\left( 
\begin{array}{cc}
T_{N-1} & \kappa I_{2^{N-1}} \\ 
\mathbf{0} & T_{N-1}%
\end{array}
\right) , \\
T_{1} &=&\left( 
\begin{array}{cc}
0 & \kappa \\ 
0 & 0%
\end{array}
\right) ,
\end{eqnarray}%
which means that after the Jordan decomposition, the size of the largest
Jordan block of $T_{N}$\ is $1$\ larger than that of $T_{N-1}$. Because $%
T_{1}$\ is a $2$-order Jordan block, the largest Jordan block\ of $h$\ is $%
\left( N+1\right) $-order, which means that the $N$-fermion system has
exceptional point of $\left( N+1\right) $-order.

\bibliographystyle{iopart-num}

\providecommand{\newblock}{}

\end{document}